# Dynamic Hotspot Migration and Self-Induced Resonances in Ultrafast Laser-Irradiated Silicon Thin Films

George D. Tsibidis[1,a)], Yi Wang[1], and Emmanuel Stratakis[1,2]

[1]*Institute of Electronic Structure and Laser (IESL), Foundation for Research and Technology (FORTH), Vassilika Vouton, 70013, Heraklion, Crete, Greece*

[2]*Department of Physics, University of Crete, 71003, Heraklion, Greece*

a) Authors to whom correspondence should be addressed: tsibidis@iesl.forth.gr

**ABSTRACT**

The spatial distribution of energy in silicon thin films under femtosecond laser irradiation is demonstrated to be influenced by the interaction between carrier-induced refractive index modulation and self-induced Fabry-Pérot resonances. We resolve three characteristic excitation regions such as interface-localized, binodal and multi-nodal regimes considering the analysis of silicon-on-insulator (SOI) configurations across thicknesses of 100, 230, and 350 nm. We report a nonlinear hotspot migration effect where the transient reduction of the refractive index stretches the internal standing wave, translating intensity antinodes toward the film volume during the pulse. Numerical results demonstrate that at higher fluences, plasma-induced extinction suppresses internal reflections, triggering a transition from resonance-mediated absorption to plasma shielding. These findings establish a physical basis for depth-resolved modification in laser nanomanufacturing, providing a dual-capacity for simultaneous surface patterning and depth-resolved subsurface modification with direct applications in laser-assisted interfacial separation and 3D integrated photonic fabrication.

Ultrafast laser modification of silicon-on-insulator (SOI) architectures is fundamental to the fabrication of high-performance microelectronics and integrated photonic devices [1-4]. While ultrafast laser processing of semiconductor surfaces has been extensively investigated, most studies have focused on the formation of Laser-Induced Periodic Surface Structures (LIPSS) on bulk materials. [5-14]. In particular, on bulk silicon, these self-organized ripples generated from the interference between the incident field and surface scattered waves or surface plasmon polaritons have become a standard for modulating surface wettability, tribology and optical absorption [8, 15, 16]. While bulk LIPSS provide a well-established framework for large-scale surface nanostructuring, the interaction is fundamentally confined to the near-surface region due to the high absorption of silicon at standard ultrafast wavelengths and the absence of internal boundary reflections. As a result, incorporating the third dimension (depth) as a design parameter remains challenging in bulk processing. On the other hand, while the energy deposition in bulk semiconductors is traditionally described by exponential decay, thin-film systems introduce a complex optical environment governed by Fabry-Pérot resonances [17, 18]. In these systems, the dielectric contrast between the high-index silicon and a low-index substrate (i.e. fused silica) creates a resonant cavity that significantly redistributes the internal electric field [19].

A critical challenge in modeling these interactions is the transient nature of the optical cavity. While energy deposition in SOI architectures is traditionally described using static interference approximations, such models fail to describe the significant, sub-picosecond reconfiguration of the resonant cavity [17, 18]. This discrepancy leads to a fundamental inability to predict the exact depth of material modification in high-precision manufacturing. To address this challenge, via the implementation of a self-consistent dynamic multireflection theory, we resolve the underlying physical mechanism that characterises the transition between resonant regimes, revealing that energy deposition is not a static state but a dynamic evolution.

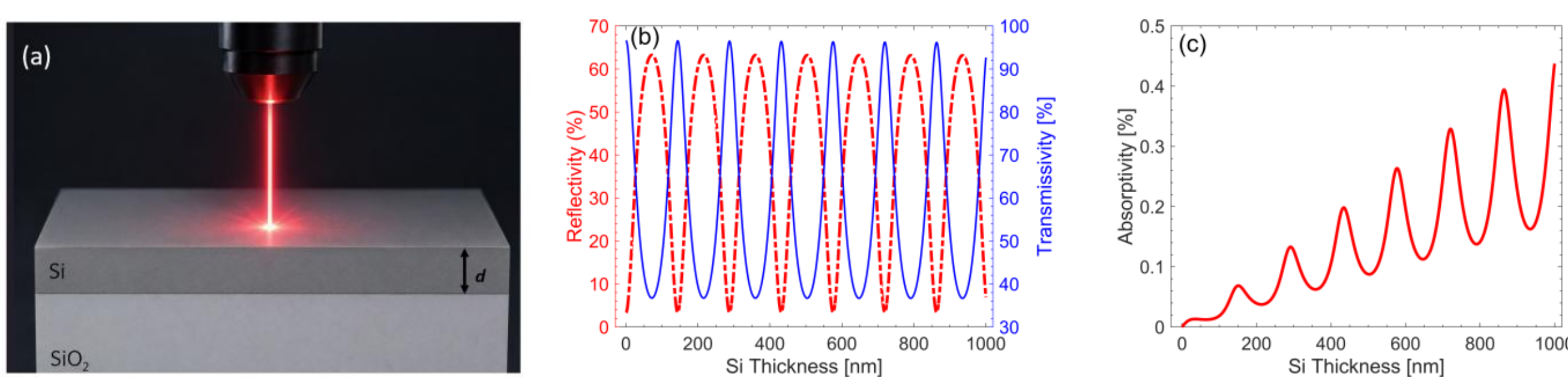

**Figure 1:** SOI's (a) Architecture, (b) Reflectivity/Transmissivity (at room temperature) and (c) Absorptivity.

This work reports on the identification of three distinct physical regimes of energy localization in Si thin films: interface enhanced (100 nm), bi-nodal (230 nm), and multi-nodal (350 nm) modes. We show that at moderate fluences ($F$=0.4 J/cm²), the zero-phase-shift reflection at the $Si/SiO_2$ interface generates intensity antinodes at the boundary between the two materials, enabling energy deposition gradients that are reversed relative to the incident direction. On the other hand, we observe a hotspot migration effect at $F$=0.7 J/cm², where the stretching of the internal wavelength (i.e. the wavelength inside the film, $\lambda_L/n_{Si}$, where $\lambda_L$ and $n_{Si}$ stand for the laser wavelength and refractive index of the material, respectively) shifts intensity peaks deep into the film volume. These findings provide a foundation for depth-resolved structuring in laser nanomanufacturing, enabling controlled subsurface modification and highly selective processing at buried interfaces.

The investigation focuses on the interaction of 170 fs laser pulses ($\lambda_L$=1026 nm) with Si layers of varying thickness $d$ (i.e. 100, 230 and 350 nm) deposited on $SiO_2$ substrates. To describe the non-equilibrium dynamics following the femtosecond irradiation of the SOI architectures (Figure 1a), we employ a self-consistent model integrating carrier-density excitation, the two-temperature model for semiconductors and dynamic multireflection theory, based on the approach detailed by Tsibidis *et* al [20]. For the sake of simplicity, films of thicknesses which are remarkably smaller than the size of the spot radius of the laser beam are considered that allows to model the multiscale processes in one dimension (along the $z$-axis). Nevertheless, results confirm that similar phenomena also occur at higher dimensions. The laser energies used in the simulations are sufficient to excite carriers in the Si layer. However, the energy transmitted into the substrate is not large enough to generate excited carriers because of both its reduced magnitude and the larger bandgap of the $SiO_2$ substrate. Thus, no significant electron excitation occurs in the substrate; by contrast, the $SiO_2$ temperature increases only slightly near the interface as heat is transferred from the Si layer through thermal diffusion. The spatio temporal distribution of the carrier density $N_e$, and the energy transport between the electron and lattice subsystems are described by a system of coupled partial differential equations [20]. It is noted that the source term and the local intensity across the irradiated material thickness are calculated using a dynamic multireflection matrix method. Unlike static models, which assume room temperature optical parameters (refractive index is equal to $N_{si}$=3.5632+$i$0.00027806 at $\lambda_L$=1026 nm [21] at 300K), the present approach correlates the variation of the refractive index with Si thickness and its impact on the optical response of the SOI structure. In static models, the calculated reflectivity and transmissivity values result in virtually zero absorption in Si, as illustrated in Figure 1b,c[18]; by contrast, in our approach, the complex refractive index $n_{Si}+ik_{Si}$ is considered to be a $N_e$-dependent variable [20] that, in turn, leads to an absorbed energy from the semiconducting film; $k_{Si}$ stands for the extinction coefficient inside the film. The model incorporates a nonlinear feedback mechanism in which the increasing carrier density reconfigures the internal standing-wave structure that subsequently controls the spatial distribution of energy deposition considering the solution of the optical and thermal equations simultaneously (see Ref. [20]),. The spatio-temporal evolution of the internal optical intensity for the varying SOI architectures is depicted in Figure 2. Results demonstrate that the spatio-temporal variation of the carrier densities (Figure 3) influence the refractive index distribution (Figure 4) which results in the remarkably different energy deposition from that predicted for bulk materials.

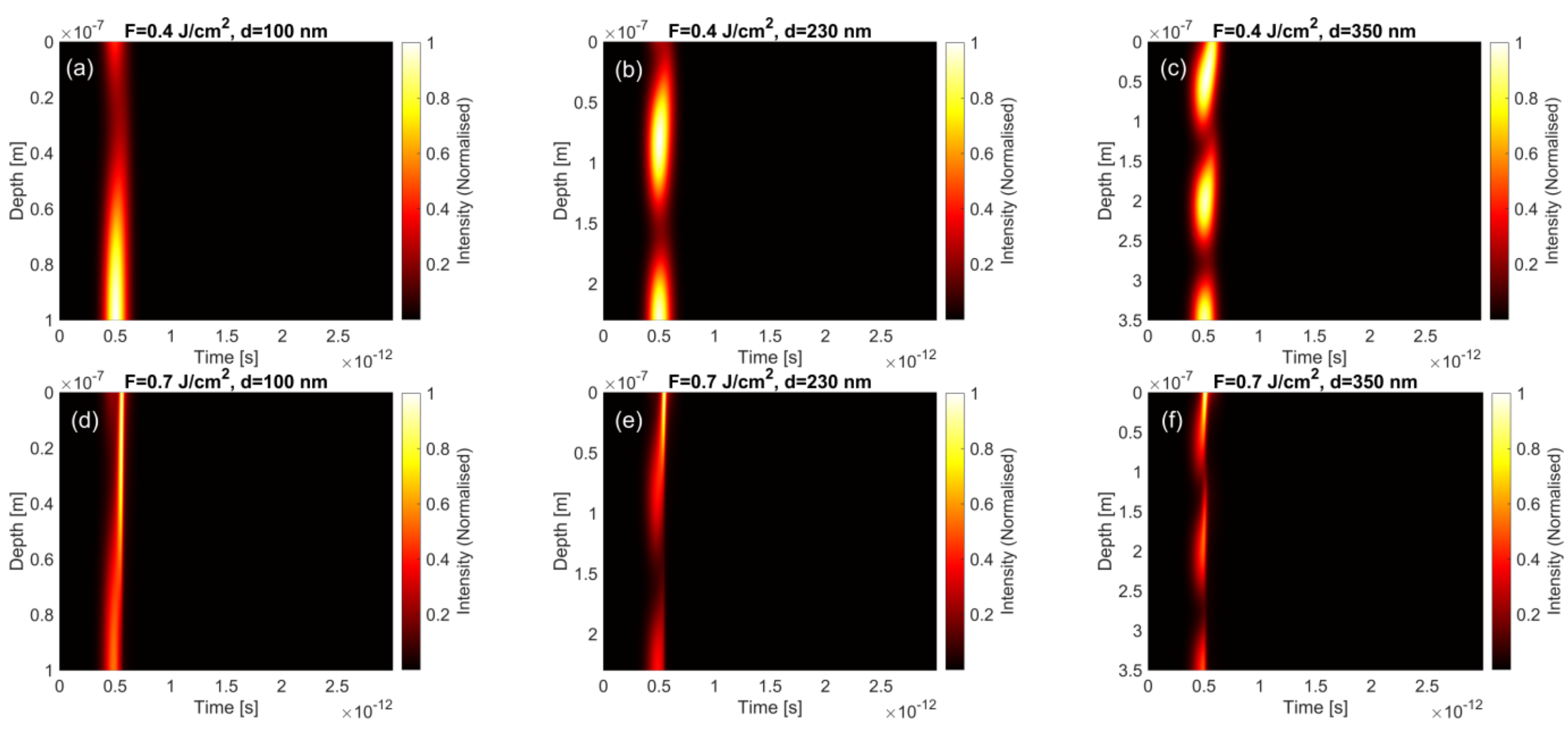

**Figure 2:** Spatio-temporal distribution of the intensity inside Si for $F$=0.4 J/cm$^2$ (first row) and $F$=0.7 J/cm$^2$ (second row) for $d$=100, 230, 350 nm.

Simulation results illustrated in Figure 2 show that at a moderate incident fluence of 0.4 J/cm² (Figures 2a-c), the energy deposition is mainly regulated by the formation of internal standing waves where the spatial intensity profile is a function of the film's cavity order $m$. Because the refractive index of the silicon remains higher than that of the fused silica substrate (i.e. $n_{Si}>n_{SiO2}$), Fresnel reflection at the $Si/SiO_2$ interface occurs with a zero-phase shift. This boundary condition enforces a displacement antinode at the $Si/SiO_2$ interface ($z=d$) which operates as the predominant factor that generates the resonant field distribution.

In regard to the 100 nm film (Figure 2a), the thickness is insufficient to accommodate an internal destructive interference node (at $\lambda_L/(4n_{Si})$~100 nm for $n_{Si}$ ~2.6) as it is comparable with the internal quarter-wavelength. This results in a monotonic intensity gradient that favours energy localization at the interface; this resembles an 'inverted' profile ideal for interface-selective applications like laser-induced forward transfer [19, 22, 23].

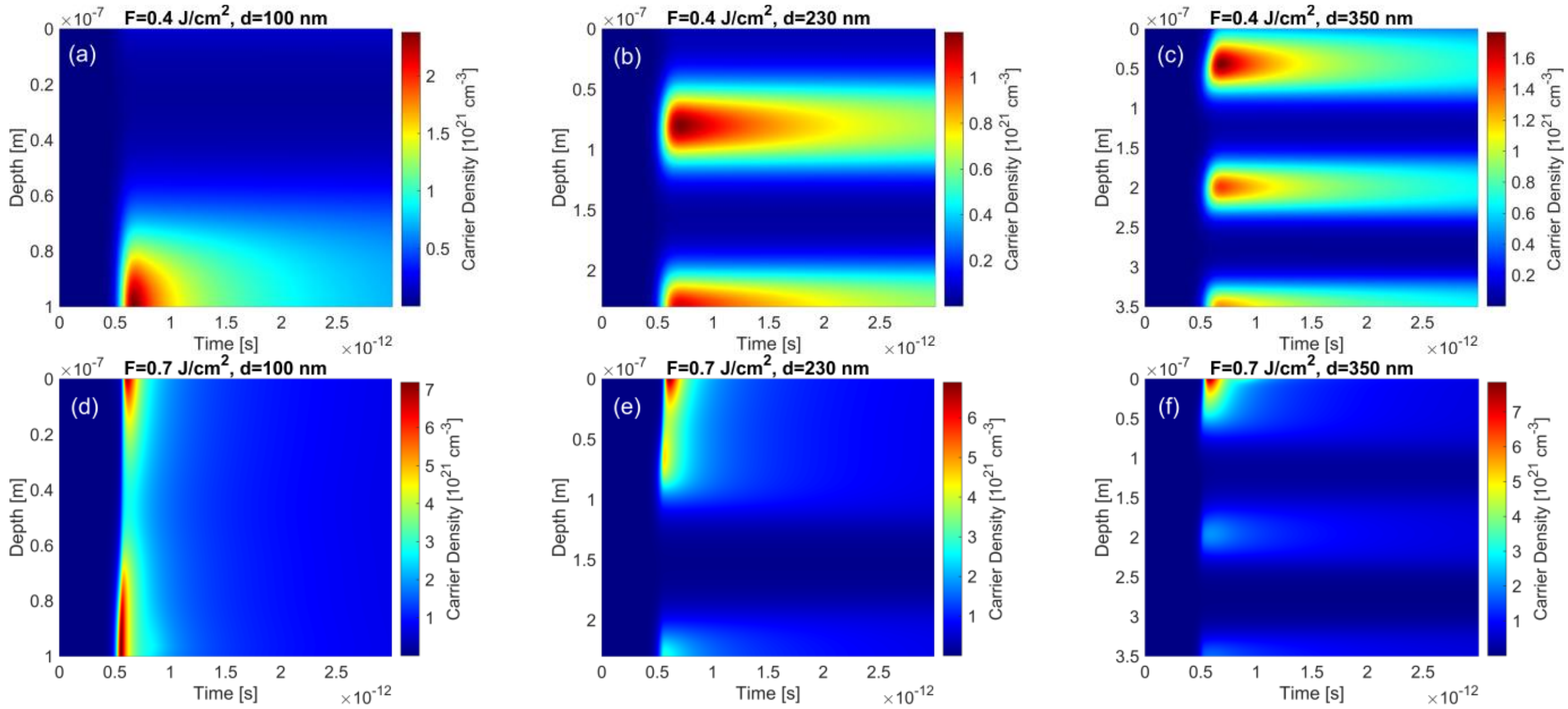

**Figure 3:** Spatio-temporal distribution of carrier density inside Si for $F$=0.4 J/cm$^2$ (first row) and $F$=0.7 J/cm$^2$ (second row) for $d$=100, 230, 350 nm.

For the 230 nm Si film at 0.4 J/cm² (Figure 2b), the energy deposition is manifested as a binodal shape, functioning as a half- wave ($\lambda_L/(2n_{Si})$) resonant cavity. The zero-phase-shift reflection exhibits a displacement antinode at the Si/$SiO_2$ boundary ($z$=230 nm). Unlike the sub-wavelength 100 nm film, the 230 nm thickness provides sufficient optical path length to accommodate a full nodal transition. As a result, a destructive interference node is resolved in the central region ($z$~155 nm), followed by a secondary antinode at $z$~84 nm. This depth-dependent energy profile generates an interesting excitation region, characterized by higher excitation intensity near the surfaces and lower intensity in the interior; it is evident that the film boundaries are preferentially modified while the central region remains protected by the standing wave node. The spatial positions of these field extrema are practically characterised by the value of the refractive index. The antinodes resolved at 0.4 J/cm² correspond to the near-static index of silicon ($n_{Si}$ ~3.5632) established at the leading edge of the pulse. The second antinode at $z$~84 nm satisfies the resonant condition $z_{max}=d-\lambda_L/(2n_{Si})$. However, as the pulse evolves, the carrier-induced refractive index collapse (reaching $n_{Si}$~2.5 at the peak) stretches the internal wavelength, translating these resonant nodes toward the surface. This dynamic reconfiguration demonstrates that the film functions as a self-tuning cavity where the pulse intensity itself precisely modulates the depth-resolved localization of the energy hotspots.

For the 350 nm Si film (Figure 2c), the system exhibits a high-order multi-nodal regime with a pronounced depth resolution. At 0.4 J/cm², the zero-phase-shift reflection at the Si/$SiO_2$ interface forces the primary antinode to be formed at the Si/$SiO_2$ boundary ($z$=350 nm). The film thickness accommodates three distinct constructive interference maxima resolved at approximately $z$~350, 203 and 57 nm consideering the near-static refractive index of $n_{Si}$~3.5632 established at the leading edge of the pulse. This periodic energy distribution effectively creates a self-induced volumetric grating within Si. As the pulse evolves, the carrier-induced decrease of the refractive index toward $n_{Si}$~2.5 at the peak intensity leads to an internal wavelength equal to 412 nm. This effect shifts the resonant nodes toward the surface, as revealed by the non-vertical temporal alignment of the intensity hotspots. Such a dynamic reconfiguration of high-order cavity modes allows for precise, multi-layered nanostructuring and depth-resolved phase modification in a single irradiation event.

For the 100 nm Si film at 0.7 J/cm² (Figure 2d), the system transitions from an interface-localized regime to a surface dominant plasma shielding regime. The rapid generation of a high-density electron-hole plasma (Figure 3d) causes the refractive index to decrease further (i.e. reaching transient values of $n_{Si}$~1.8) while simultaneously producing a sharp rise in the extinction coefficient $k_{Si}$. This large refractive index reduction results in a significant increase of the internal wavelength to $\lambda_L/n_{Si}$~570 nm. Under these conditions, the distance required to resolve a destructive interference node ($\lambda_L/(4n_{Si})$ increases to approximately 143 nm. Since the film thickness ($d$=100 nm) remains well below this resonant threshold, the first node is practically shifted further outside the film volume at $z=d-\lambda_L/(4n_{Si})$ (~-43 nm). While a displacement antinode remains present at the Si/$SiO_2$ interface due to the zero-phase-shift boundary condition, the high internal extinction reduces the reflected wave, decreasing the resonant contrast. Thus, the intensity profile becomes more uniform and shifts towards the Si surface ($z$=0 nm) where the absorption of the high- density plasma acts as an optical shield for the underlying substrate.

For the 230 nm Si film at 0.7 J/cm² (Figure 2e), the system exhibits a pronounced nonlinear hotspot migration and a transition toward a surface-shielded regime. An enhanced carrier generation causes the refractive index to decrease more substantially, approaching transient-state levels of $n_{Si}$ ~2-2.2, which stretches the internal standing wave pattern. This refractive index reduction increases the internal wavelength to ~ 515 nm, effectively shifting the resonant cavity conditions. The destructive interference node, previously centred at $z$~155nm at lower fluences, is moving toward the surface as the quarter-wavelength threshold ($\lambda_L/4$) increases to ~129 nm. Thus, the second antinode (resolved at $z$~84 nm at the lower fluence) is shifted toward the surface of the Si ($z$=0). This suppresses the field reflected from the substrate, leading to a diminished nodal contrast and a reduced interface-localized intensity at $z$=230 nm. The above results indicates that, at high fluences, the cavity's self-adjusting behaviour promotes energy deposition near the front surface while inhibiting the development of sub-surface volumetric resonances.

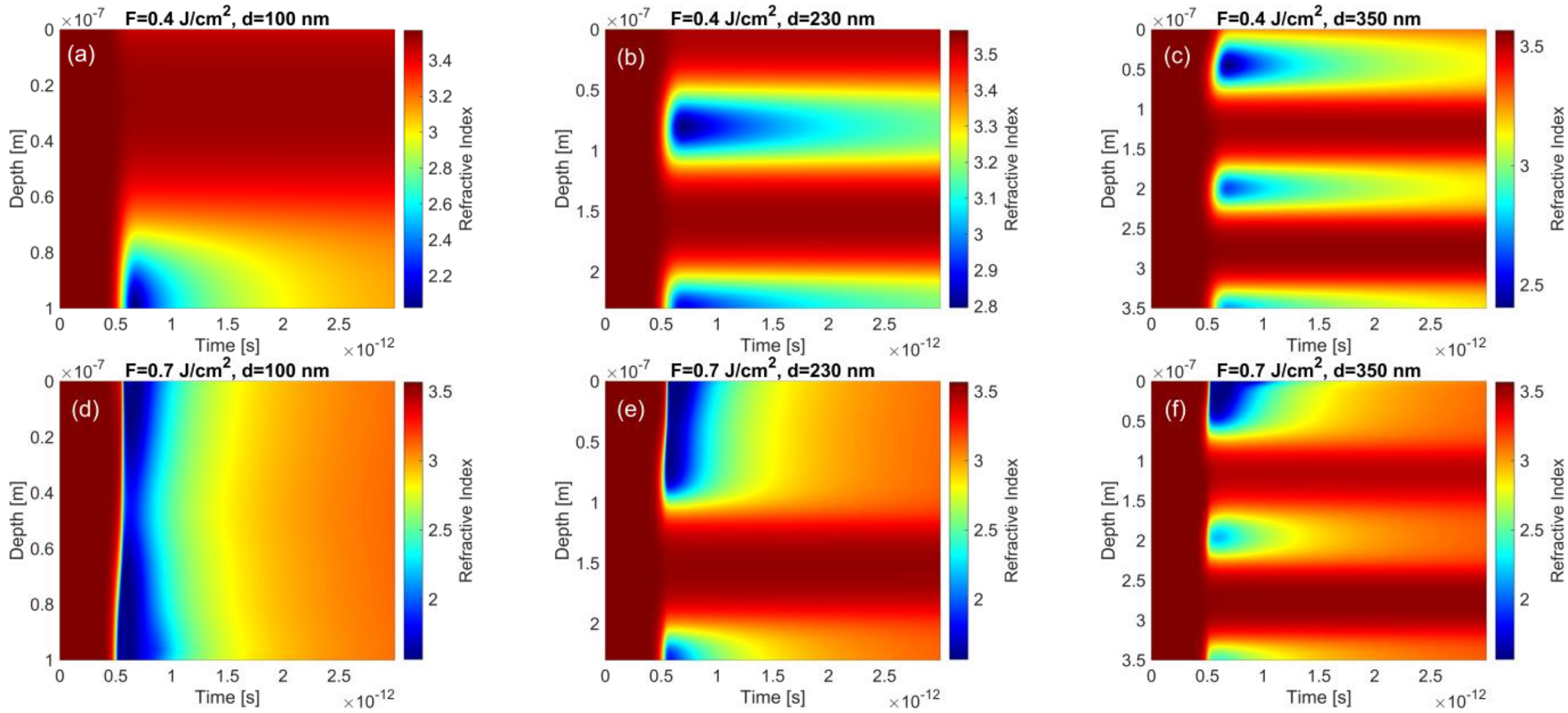

**Figure 4:** Spatio-temporal distribution of refractive Index for $F$=0.4 0.7 J/cm$^2$ (first row) and $F$=0.7 J/cm$^2$ (second row) for $d$=100, 230, 350 nm.

Finally, for the 350 nm Si film at 0.7 J/cm² (Figure 2f), the high-order resonant regime is fundamentally altered by extinction-induced reduction and nonlinear hotspot migration. The enhanced photoexcitation leads to a large decrease of the refractive index toward transient values of $n_{\mathrm{Si}}$ ~1.9-2.1, while simultaneously triggering a remarkable rise in the internal extinction coefficient. The reduction in the refractive index physically stretches the internal standing wave as the internal wavelength expands to ~515 nm. In these conditions, the energy that is absorbed from Si exhibits the following regimes: the second antinode resolved at $z$~203 nm at lower fluences, is shifted toward the surface to a depth of $z$~93nm. Furthermore, the third antinode which was located near the surface is shifted out of the film volume ($z$<0). Most notably, the high carrier density ($N_e$~7×10$^{21}$ cm$^{-3}$, Figure 3) induces a suppression of the high-order interference fringes. The increased internal extinction attenuates the laser field as it propagates toward the substrate, reducing the amplitude of the reflected wave from the Si/$SiO_2$ interface. Thus, the multi-nodal contrast is suppressed, and the energy deposition profile transitions from a volumetric grating to a surface-limited plasma shielding regime. The primary energy deposition is confined to the front surface, where the dense plasma layer acts as an optical shield, significantly reducing the intensity of the interface-situated antinode at $z$=350 nm. This transition defines the operational limit for resonant volumetric structuring where energy absorption-driven reduction dominates over interference effects. By contrast, a static calculation based on the initial refractive index of silicon ($n_{\mathrm{Si}}$ ~3.5632) would predict stationary intensity maxima. Thus, our results demonstrate a previously unobserved nonlinear hotspot migration. As the carrier-induced index reduction stretches the internal wavelength, the intensity antinodes physically translate toward the surface. This self-tuning behaviour of the cavity defines the operational limits for precise depth-resolved structuring.

Figure 3 illustrates a detailed spatio-temporal evolution of carrier density, where at 0.4 J/cm² the profiles demonstrate the resonant intensity antinodes established by the cavity modes. At the higher fluence of 0.7 J/cm², rapid photoexcitation generates a high-density plasma (reaching ∼7×10$^{21}$ cm$^{-3}$) that increasingly localizes at the front surface. This saturation of $N_e$ triggers the transition toward a plasma shielding regime, which effectively attenuates the laser field and prevents significant energy deposition at the Si/$SiO_2$ interface. On the other hand, Figure 4 depicts the spatiotemporal reduction of $n_{\mathrm{Si}}$, which drops from its near-static value of 3.5632 to as low as ~1.8 during the pulse peak at high fluences. This carrier induced modulation physically stretches the internal wavelength, resulting into a dynamical reconfiguration of the cavity on a sub-picosecond timescale. This reduction in $n_{\mathrm{Si}}$ is the fundamental reason of the reported nonlinear hotspot migration, as it physically translates the intensity antinodes toward the surface as the pulse evolves.

The spatio-temporal evolution of the lattice temperature depicted in Figure 5, illustrates how the energy localized within the self-induced resonant modes is transferred to Si film via electron-phonon coupling. At the moderate fluence of 0.4 J/cm² (Figures 5a-c), the thermal distribution provides a high-precision characterization of the internal standing-wave patterns obtained from optical intensity simulations. For the 100 nm architecture, the thermal fingerprint is exclusively interface-localized, with temperatures reaching values ∼1600 K at the Si/$SiO_2$ boundary, confirming that sub-wavelength films can facilitate selective substrate heating while preserving surface integrity. As the thickness increases to 230 nm, a binodal thermal shape emerges, characterized by distinct heating zones at the film boundaries separated by a persistent central thermal well at $z$~150 nm, where temperatures remain close to ambient. This depth-dependent partitioning is further extended in the 350 nm film, which manifests a volumetric thermal grating composed of three distinct heating layers. By contrast, the transition to the 0.7 J/cm$^2$ regime (Figures 5d-f) exhibits a fundamentally different configuration of the thermal energy of the lattice system, dictated by the onset of plasma shielding and nonlinear hotspot migration. In the 100 nm case, the lattice reaches extremely high temperatures exceeding 6000 K at both the interface and the surface, reflecting the convergence of resonant trapping and absorption of the Si surface. For the 230 nm architecture, the sub-surface migration of the intensity antinode is confirmed by the displacement of the thermal maximum into the film volume, while the interface heating is suppressed. Interestingly, the multi-nodal field distribution of the 350 nm film effectively disappears in this high-fluence regime; the carrier-induced increase in the extinction coefficient confines the lattice heating to the first ~50 nm of the film, where temperatures surpass 8000 K. This confirms that at high excitation levels, absorption-driven reduction dominates over interference effects, shifting the thermal effects toward a surface-limited ablation regime. It is emphasized that throughout these interactions, the fused silica substrate functions as a thermal insulator, preventing energy dissipation into the bulk and enabling the precise depth-resolved localization of these high-contrast resonant thermal profiles.

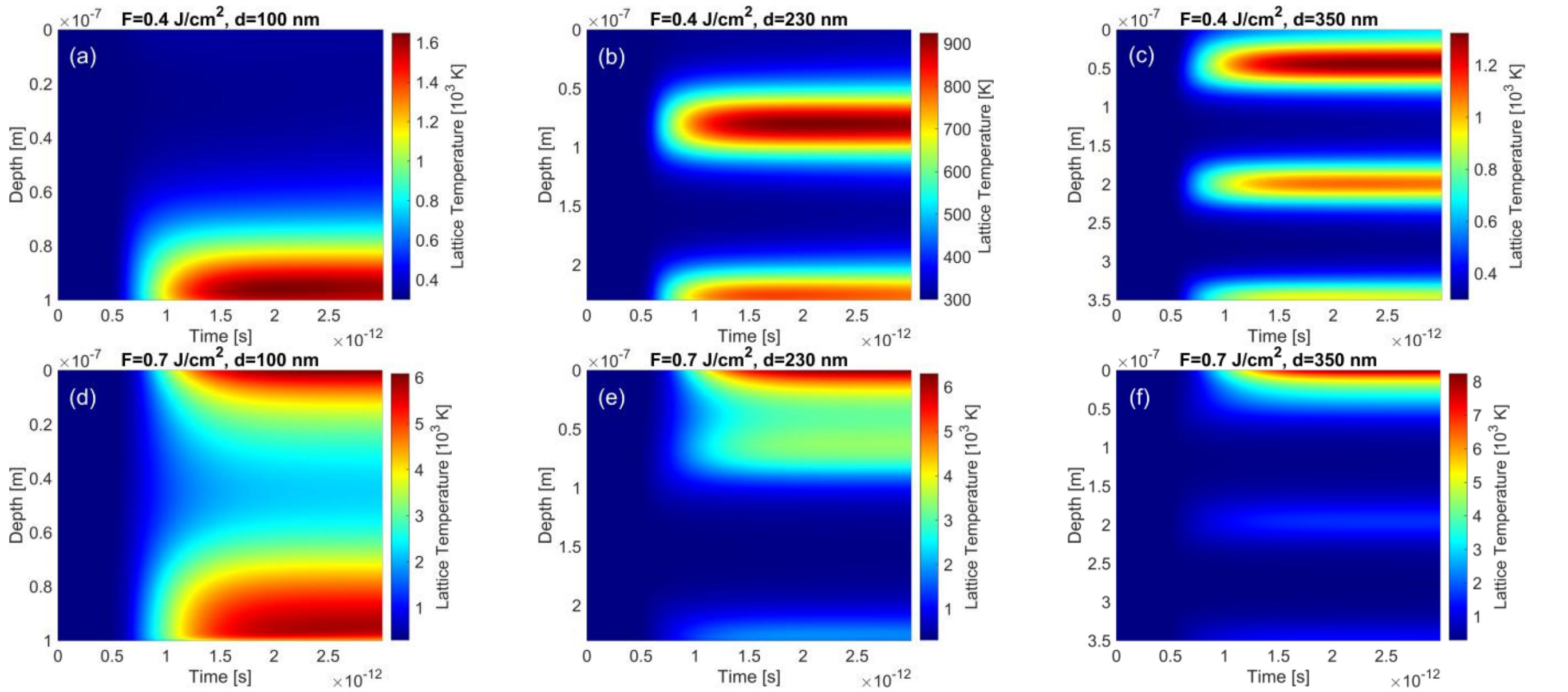

**Figure 5:** Spatio-temporal distribution of lattice temperature for *F*=0.4 J/cm$^2$ (first row) and *F*=0.7 J/cm$^2$ (second row) for *d*=100, 230, 350 nm.

In conclusion, the resolved resonant regimes establish a physical basis for precise energy deposition along the depth of the film, providing a unique dual-capacity for simultaneous surface patterning [10] and subsurface structuring. The interface-localized mode observed in 100 nm films is particularly advantageous for high-precision laser-induced forward transfer and selective delamination where energy is concentrated at the Si/$SiO_2$ boundary while avoiding the functional surface from thermal damage. In thicker architectures (230 and 350 nm), the thermally induced layered profiles enable single-shot fabrication of multi-layered structures via localized phase modification. More specifically, the periodic amorphization of resonant antinodes allows for the creation of alternating amorphous/crystalline stacks if conditions are sufficient to result into a phase transition (i.e. melting). Such depth-dependent phase changes are essential for high-contrast optical memory and the development of integrated biosensors [24], where the resulting refractive index modulation or selective etching of amorphous layers can be used to enhance sensitivity. Furthermore, this model demonstrates that by modulating the incident fluence, a transition occurs between a resonant-volumetric regime (capable of fabricating optical elements and nanofluidic channels at subsurface locations within the film) and a shielded-surface regime where plasma shielding confines modification to the air-exposed interface. This ability to direct energy deposition to any desired depth provides an advanced degree of freedom for the maskless 3D manufacturing of integrated electronic, photonic, and sensing devices, effectively bridging the gap between traditional surface-bound LIPSS and true volumetric material engineering.

This research work was supported from Horizon Europe, the European Union's Framework Programme for Research and Innovation, under Grant Agreement No. 101057457 (METAMORPHA) and European Projects Lasers4EU (Grant agreement ID: 101131771) and RIANA (Grant agreement ID: 101130652).

## AUTHOR DECLARATIONS

### Conflict of interest

The authors declare no conflicts of interest.

### Author Contributions

**George D. Tsibidis**: conceptualization of the work; physical modelling simulations (Electromagnetic 1D, excitation and thermal effects) and interpretation of the results; writing-original draft; **Yi Wang**: simulations (Electromagnetic 2D); **Emmanuel Stratakis**: reviewing-original draft; funding.